\documentstyle[aps]{revtex}
\input epsf
\newcommand{\un}{u^{(n)}}     
\newcommand{\vn}{v^{(n)}}
\newcommand{\rhon}{\rho^{(n)}} 
\newcommand{\psin}{\psi^{(n)}} 
 
\newcommand{\unmean}{\langle u^{(n)} \rangle} 
\newcommand{\vnmean}{\langle v^{(n)} \rangle}

\begin{document}
\draft
\title{Self-organization in nonlinear wave turbulence} 
 
\author{Richard Jordan$^{1,3}$ and Christophe Josserand$^{2,3}$}
  
\address{$1$ T-7, Los Alamos National Laboratory, Los Alamos, NM 87545\\ 
$2$ The James Franck Institute, University of 
Chicago, 5640 South Ellis, Chicago, Illinois 60637\\ 
$3$ CNLS, Los Alamos National Laboratory, Los Alamos, NM 87545} 

\date{\today}
\maketitle 

\begin{abstract} We present 
a statistical equilibrium model of self-organization in 
 a class of focusing, nonintegrable nonlinear Schr\"{o}dinger (NLS)
equations. The theory predicts that the asymptotic--time
behavior of the NLS system 
is characterized by the formation and persistence of
a large--scale coherent solitary wave, which minimizes
the Hamiltonian given the conserved particle number ($L^2$-norm
squared),
coupled with small--scale random fluctuations, or radiation.  The
fluctuations account for the
difference between the conserved value of the Hamiltonian and the
Hamiltonian of the coherent state.  
The predictions of the statistical theory are tested against the
results of direct numerical simulations of NLS, and excellent
qualitative and quantitative agreement is demonstrated. In addition,
a careful inspection of
the numerical simulations reveals interesting features of the
transitory dynamics leading up to the
to the long--time statistical equilibrium 
state starting from a given initial condition. As time increases,
the system investigates smaller and smaller scales, and  it
appears that at a given intermediate time after the coalescense
of the soliton structures has ended,, the system is nearly in
statistical equilibrium  over
the modes that it has investigated up to that time.
\end{abstract}


 \vspace{1.5ex} 

Submitted to {\em Physical Review E}
 
\vspace{1.5ex}

\section{Introduction: NLS and Soliton Turbulence} 
A fascinating feature of many turbulent fluid and plasma systems is 
the emergence and persistence of large--scale organized states, or 
coherent structures, in the midst of small--scale turbulent 
fluctuations. A familiar example is the formation of  
macroscopic quasi--steady vortices in a turbulent 
large Reynolds number two dimensional fluid\cite{tab,vor,Mcwilliams}. 
Such phenomena also occur for many classical Hamiltonian 
systems, even though the dynamics of these systems
 is formally reversible\cite{hasa}. 
 In the present work, we shall focus our  
attention on another class of nonlinear partial differential equations  
whose solutions exhibit the tendency to form persistent coherent  
structures immersed in a sea of microscopic turbulent fluctuations.  
This is the class of nonlinear wave systems described by the 
well--known nonlinear Schr\"odinger (NLS) equation:

\begin{equation} \label{nls1} 
i \partial_t \psi + \Delta \psi + f(| \psi |^2)\psi\ = 0\,, 
\end{equation} 
where $\psi({\bf r},t)$ is a complex field and $\Delta$ is the 
Laplacian operator. The NLS equation describes the slowly-varying  
envelope of a wave train in a dispersive conservative system. 
It models, among other things, gravity waves on deep water \cite{AS},
Langmuir 
waves in plasmas \cite{Pesceli}, pulse propagation along optical fibers 
\cite{HK}, and superfluid dynamics\cite{GP}. 
When $f(|\psi|^2)= \pm |\psi|^2$ and eqn.~(\ref{nls1}) is posed on the 
whole real line or on a bounded interval with periodic boundary 
conditions, the equation is completely integrable 
\cite{ZS}. Otherwise, it is nonintegrable.  
 
The NLS equation (\ref{nls1}) may be cast in the Hamiltonian form $i 
\partial_t \psi 
= \delta H/ \delta \psi^*$, where $\psi^*$ is the complex conjugate of the 
field $\psi$, and 
$H$ is the Hamiltonian: 
\begin{equation} \label{ham1} 
H(\psi) = \int\left ( |{\bf \nabla} \psi|^2-F(|\psi|^2) \right 
)\,d{\bf r}\,. 
\end{equation} 
Here, the {\em potential} $F$ is defined via the relation $F(a) = \int_0^a 
f(y)\, dy$. 
The dynamics (\ref{nls1}) conserves, in addition to the Hamiltonian, the 
particle number 
 
\begin{equation} \label{part1} 
N(\psi)= \int |\psi|^2\, d{\bf r}\,. 
\end{equation} 
 
We shall assume throughout that eqn.~(\ref{nls1}) is posed in a 
bounded one dimensional interval with either periodic or homogeneous  
Dirichlet boundary conditions. We restrict our attention to 
attractive, or focusing, nonlinearities $f (f(a) \ge 0,\; f'(a) >0)$ 
such that the dynamics described by (\ref{nls1}) is  
nonintegrable, free of wave collapse, and admits stable solitary--wave  
solutions. The dynamics under these conditions has been 
referred to as {\em soliton turbulence} \cite{ZPSY}. 
Such is the case for the important power law nonlinearities,  
$f(|\psi|^2)=|\psi|^s$, with  
$0 < s<4$ (in the periodic case, $s\neq 2$ for nonintegrability)  
\cite{Bourgain,RR}, and also for the physically relevant 
saturated nonlinearities $f(|\psi|^2) = |\psi|^2/(1 + |\psi|^2)$ 
and $f(|\psi|^2)= 1-\exp(-|\psi|^2)$, which arise as corrections 
to the cubic nonlinearity for large wave amplitudes \cite{Max}. 

Equation (\ref{nls1}) in one spatial dimension has
solitary wave  solutions of the form  
$\psi(x,t)=\phi(x)\exp(i \lambda^2 t)$, where $\phi$ satisfies the  
 nonlinear eigenvalue equation: 
 
\begin{equation} \label{gse} 
\phi_{xx} + f(|\phi|^2)\phi - \lambda^2 \phi = 0\,. 
\end{equation} 
It has been argued \cite{ZPSY,Pomeau} that the solitary 
wave solutions play a prominent role in the long--time  
dynamics of (\ref{nls1}), in that they act as {\em statistical  
attractors} to which the system relaxes. The numerical simulations  
in \cite{ZPSY}, as well as the simulations we shall 
present within this article, support this conclusion. 
Indeed, it is seen that for rather generic initial conditions the field 
$\psi$ evolves, after a sufficiently long time, into a state consisting of 
a spatially localized coherent structure, which compares quite 
favorably to a solution of (\ref{gse}), immersed in a sea of turbulent  
small-scale turbulent fluctuations. 
At intermediate times the solution typically consists of a collection of 
these soliton-like structures, but as time evolves, the solitons undergo a 
succession of collisions in which the smaller soliton 
decreases in amplitude, while the larger one increases in amplitude. 
When solitons collide or interact, they shed radiation,  
or small--scale fluctuations. The interaction of the solitons continues 
until eventually a single soliton of large amplitude survives amidst 
the turbulent background radiation.   
Figure (\ref{evolu}) below illustrates the evolution of the solution
of (\ref{nls1}) for the particular nonlinearity $f(|\psi|^2) = |\psi|$ 
and with periodic boundary conditions on the
 spatial interval $[0,256]$.
 
\begin{figure} 
\centerline{  \epsfxsize=10truecm \epsfbox{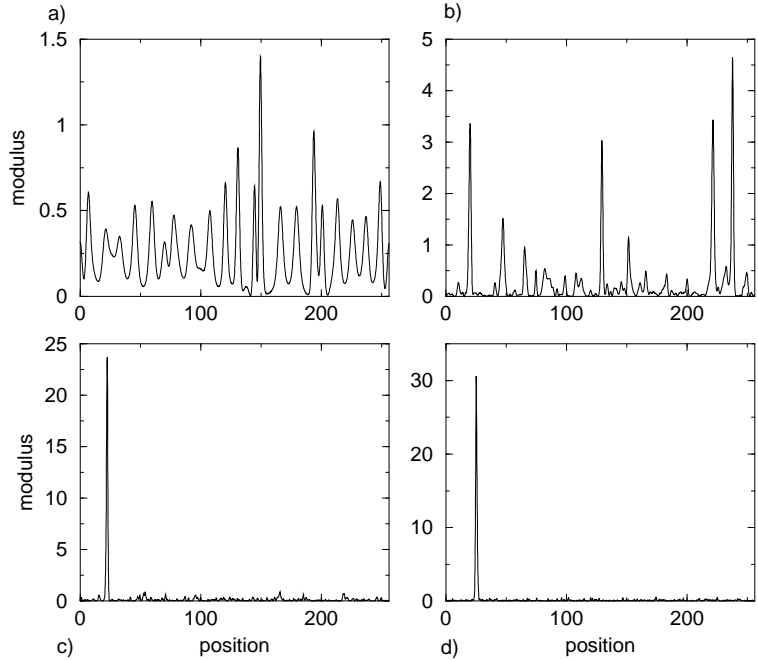}} 
\caption{\protect\small Profile of the modulus 
$|\psi|^2$ at four different times for the system (\ref{nls1})
with nonlinearity $f(|\psi|^2) = |\psi|$ and periodic boundary 
conditions on the
interval
$[0,256]$.   The initial 
condition is  $\psi(x, t=0 ) = A$, with $A = 0.5$,  plus
a small random perturbation.  The numerical scheme used to approximate
the solution is the split-step Fourier method. 
The  grid size  is $dx=0.125$, and the number of  modes is $n=2048$. 
a) $t=50$ unit time: Due to the modulational instability, an array of 
soliton-like structures separated by the typical distance  
$l_i=2\pi/\sqrt{A/2}=4\pi$ is created; b) $t=1050$ unit time: 
The solitons interact and coalesce, giving rise to a smaller number  
of solitons of larger amplitude; c)  $t=15050$:  The  
coarsening process has ended.  One large soliton  
remains in a background of  small--amplitude radiation. Notice that for  
$t=55050$ unit time (Figure d)), the amplitude of the fluctuations  
has diminished while the amplitude of the soliton has increased. 
\label{evolu}} 
\end{figure} 
In modeling the long--time behavior of a Hamiltonian system such as 
NLS, it seems natural to appeal to the methods of equilibrium 
statistical mechanics.  That such an approach may be relevant for 
understanding the asymptotic--time state for NLS 
 has already been suggested in \cite{ZPSY}, although 
the thermodynamic arguments presented by these authors are rather 
formal and somewhat incomplete. Motivated in part by the ideas  
outlined in\cite{ZPSY}, Jordan {\em et al.} \cite{JTZ} have  
recently constructed a mean--field statistical theory to characterize  
the large--scale structure and the statistics of the 
small--scale fluctuations inherent in the asymptotic--time state of 
the focusing nonintegrable NLS system (\ref{nls1}).  The main prediction
of  
this theory is that the coherent state that emerges in the long--time 
limit is the ground state solution of equation (\ref{gse}). That is, 
it is the solitary wave that minimizes the Hamiltonian $H$ given the  
constraint $N=N^0$, where $N^0$ is the initial and conserved value of the 
particle number integral. This prediction is in accord with 
previous theories\cite{ZPSY,Pomeau}, but the approach taken in 
\cite{JTZ} is new, and provides a definite interpretation to the 
notion set forth in the earlier works that it is  
``thermodynamically advantageous'' for the NLS system to approach 
a coherent solitary wave structure that minimizes the Hamiltonian 
subject to fixed particle number. The statistical theory also gives  
predictions for the particle number spectral density and the kinetic  
energy spectral density, at least for a finite--dimensional 
spectral truncation of the NLS dynamics (\ref{nls1}). 
In particular, it predicts an equipartition of kinetic energy among 
 the small--scale fluctuations. 
 
In the present work, we shall begin with a brief 
review of this statistical theory. The predictions of the  
statistical theory will  then 
be compared in detail with the results of direct numerical simulations of the
NLS system. 
In addition, we will also closely examine the 
evolution of the particle number spectrum in our
numerical simulations, as well as the dynamics  (of finite spectral
approximations) of the integrals
$ S_m(\psi) =\int |D^m \psi|^2\, dx$ (here $D^m$ denotes the $m$--th
derivative
with respect to the spatial variable). The statistical model, being  
strictly an equilibrium theory, does not give predictions concerning 
the finite time dynamics of these quantities. However, we shall see that
it does give accurate estimates for the long--time saturation values
of these quantities for a finite dimensional spectral approximation of
the NLS dynamics. In addition, we will demonstrate that the integrals
$S_m$ exhibit interesting power law growth in time, as suggested 
by the weak turbulence theory developed by Pomeau\cite{Pomeau}.

\section{Mean-Field Statistical Model} 
 
In order to develop a meaningful statistical theory, we begin by
introducing a 
finite--dimensional approximation of the NLS equation (\ref{nls1}). 
To fix ideas and notation, we will consider the NLS system 
with homogeneous Dirichlet boundary conditions on an interval $\Omega$ 
of length $L$. Our methods can easily be modified to accommodate
other boundary conditions, and we will consider below the predictions
of the theory for periodic boundary conditions, as well. In addition, our 
techniques can easily be extended to higher dimensions, but we wish  
to concentrate on the one--dimensional case for ease of presentation.  
 
Let $e_j(x)=\sqrt{2/L}\sin(k_j x)$ with $k_j=\pi j/L$, and for any 
function $g(x)$ on $\Omega$ denote by $g_j = \int_{\Omega} g(x)e_j(x)\,
dx$ 
its $j$th Fourier coefficient with respect to the orthonormal basis $e_j, 
 j= 1,2,\cdots$.  Define the functions 
$u^{(n)}(x,t) = \sum_{j=1}^n u_j(t) e_j(x)$ and 
$v^{(n)}(x,t) = \sum_{j=1}^n v_j(t) e_j(x)$, where the real 
 coefficients $u_j,v_j, j=1, \cdots, n$, satisfy the coupled system of  
ordinary differential equations 
 
\begin{equation} 
\begin{array}{c} 
\dot{u}_j - k_j^2 v_j +  
\left ( f( (u^{(n)})^2 + (v^{(n)})^2) v^{(n)} \right )_j = 0\\ 
\\ 
\dot{v}_j + k_j^2 u_j -  
\left ( f( (u^{(n)})^2 + (v^{(n)})^2) u^{(n)} \right )_j = 0\,. 
\end{array} 
\label{spectrunc} 
\end{equation} 
Then the complex function $\psin = \un + i\vn$ satisfies the equation 

\begin{displaymath}
 i \psin_t + \psin_{xx} + P^n ( f(|\psin|^2) \psin ) = 0\,,
\end{displaymath}
 where 
$P^n$ is the projection onto the span of the eigenfunctions 
$e_1, \cdots, e_n$. This equation is a natural spectral approximation 
of the NLS equation (\ref{nls1}), and it may be shown that its 
solutions  converge as $n \rightarrow \infty$ to solutions of 
(\ref{nls1}) \cite{Bourgain,Zhidkov}. 
 
For given $n$, the system of  equations (\ref{spectrunc})  
defines a dynamics on the $2n$--dimensional phase space ${\bf R}^{2n}$.
This  
finite-dimensional dynamical system is a Hamiltonian system, with  
conjugate variables $u_j$ and $v_j$, and with Hamiltonian 
 
\begin{equation} \label{disham} 
H_n = K_n + \Theta_n\,, 
\end{equation} 
where 
\begin{equation}\label{kn} 
K_n = \frac{1}{2} \int_{\Omega} ((u^{(n)}_x)^2 + (v^{(n)}_x)^2)\, dx = 
 \frac{1}{2} \sum_{j=1}^n k_j^2 (u_j^2 + v_j^2)\,, 
\end{equation} 
is the kinetic energy, and 
\begin{equation} \label{tn} 
\Theta_n = -\frac{1}{2}\int_{\Omega} F((u^{(n)})^2 + (v^{(n)})^2)\\ 
dx\,, 
\end{equation}  
is the potential energy. 
The Hamiltonian $H_n$ is, of course, an invariant of the dynamics.  
The truncated version of the particle number 
 
\begin{equation}\label{dispn} 
N_n = \frac{1}{2} \int_{\Omega} ( (u^{(n)})^2 + (v^{(n)})^2)\, dx = 
 \frac{1}{2} \sum_{j=1}^n (u_j^2 + v_j^2)\,, 
\end{equation} 
is also conserved by the dynamics (\ref{spectrunc}). The factor  
$1/2$ is included in the definition of the particle number for
convenience. 
The Hamiltonian system (\ref{spectrunc}) satisfies the 
Liouville property, which is to say that 
the measure $\prod_{j=1}^n du_j dv_j$ is invariant under the dynamics 
\cite{Bid}. This property together with the assumption 
of ergodicity of the dynamics provide the usual starting point for a 
statistical treatment of a Hamiltonian system\cite{Balescu}. 

With the finite dimensional Hamiltonian system in hand, we now 
consider a macroscopic description in terms of a probability density  
$\rhon(u_1, \cdots, u_n, v_1\,\cdots, v_n)$  
on the $2n$--dimensional phase--space 
${\bf{R}}^{2n}$.  We seek a probability density that describes the 
statistical equilibrium state for the truncated dynamics.  In accord 
with standard statistical mechanics and information theoretic 
principles, we define this state to be the density $\rhon$ on 
$2n$--dimensional phase space which 
maximizes the Gibbs--Boltzmann entropy functional 
 
\begin{equation}\label{gbent} 
S(\rho) = -\int_{{\bf{R}}^{2n}} \rho \log \rho  
\prod_{j=1}^n du_j dv_j\,, 
\end{equation} 
subject to constraints dictated by the conservation of the 
Hamiltonian and the particle number under the dynamics  
(\ref{spectrunc}) \cite{Balescu,Jaynes}.  
 
\begin{figure} 
\centerline{  \epsfxsize=10truecm \epsfbox{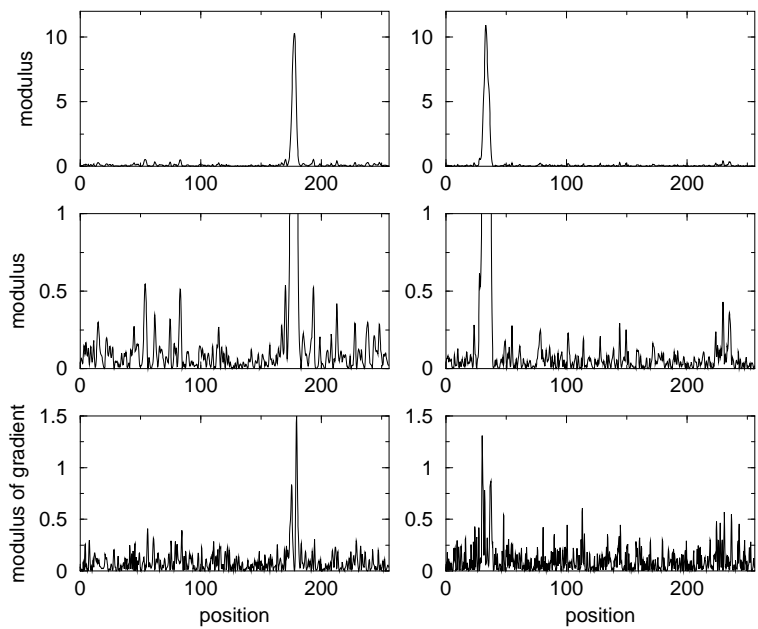}} 
\caption{\protect\small Numerical simulation for the saturated 
non-linearity $ f(|\psi|^2)=|\psi|^2/(1+|\psi|^2)$ and for periodic
boundary conditions.  The total number of modes is $n=1024$
and the spatial grid size is $dx=0.25$, so that the length of periodic
interval is $L=256$. Displayed are the 
modulus of the field $|\psi|^2$ (first and second rows),
and the modulus of the
gradient of the field $|\psi_x|^2$ (third row) at unit times
$t=30,000$
(left) and $t=220,000$ (right).  The second row shows the same results
as the first row, except that the we have restricted the range on the 
vertical axis in order to focus in on the the fluctuations of the field.
Notice that the dynamics for this saturated nonlinearity is
qualitatively similar to that for the power law nonlinearity 
$f(|\psi|^2) = |\psi|$ shown in Fig.~(\ref{evolu}):  the long--time
state consists
of large-scale coherent solitary wave--like structure interacting with a
sea of
small-scale fluctuations (top row). The typical amplitude of the 
fluctuations of the
field has decreased from $t=30,000$ to $t=220,000$ (second row), while
 the amplitude of the coherent structure has increased somewhat. 
The maximum of the modulus of the field is on the order of
 50 times larger than the typical modulus of the fluctuations
 at $t=220,000$.
On the other hand, the typical amplitude of the fluctuations of the
gradient
of the field has actually increased somewhat 
from $t=30,000$ to $t=220,000$, and the typical amplitude of the
fluctuations
of the gradient is only several times smaller than the maximum
amplitude of the gradient of the field (bottom row).  Clearly, the fluctuations
make a significant contribution to the kinetic energy in the long-time
limit.
\label{grad}} 
\end{figure}

The usual canonical ensemble 
\begin{displaymath}
\rho \propto
\exp \left ( -\beta H_n - \mu N_n \right )\,,
\end{displaymath}
results from maximizing the entropy
subject to the mean constraints $\langle H_n \rangle=H^0$ and  
$\langle N_n \rangle=N^0$, where $H^0$ and $N^0$ are the given values of
the Hamiltonian and the particle number, respectively, and
$\beta$ and $\mu$ are the Lagrange multipliers to enforce these
constraints. However, it has been shown in
\cite{JTZ,LRS} that, for the focusing nonlinearities we consider
here, the canonical ensemble
is ill--defined in the sense that it is not normalizable (i.e.,
$\int_{{\bf R}^{2n}} \exp [ -\beta H_n - \mu N_n ]
\prod_{j=1}^n du_j dv_j$ diverges ).
Thus, we are obliged to consider an alternative statistical
equilibrium description of the NLS system based
on constraints other than those that give rise to the canonical
ensemble.
The key to constructing 
an appropriate statistical model is
based on the observation from numerical simulations 
that, for a large number of modes $n$, in the 
long--time limit, the field $(u^{(n)}, v^{(n)})$ decomposes into two  
essentially distinct components: a large--scale  coherent 
structure, and small--scale radiation, or fluctuations. As time 
progresses, the amplitude of the fluctuations decreases, until 
eventually the contribution of the fluctuations to the particle 
number and the potential energy component of the Hamiltonian becomes 
negligible compared to the contribution from the coherent state, so 
that $N_n$ and $\Theta_n$ are determined almost 
entirely by the coherent structure. We have checked that
 this effect becomes even more 
pronounced when the resolution of the numerical simulations is improved 
(i.e., when the number of modes is increased with the length
$L$ of the spatial interval fixed).  On the other 
hand, as the fluctuations  exhibit rapid  spatial variations, 
the amplitude of their gradient does not, in general, become 
 negligible in the asymptotic time limit. Consequently,  
the fluctuations can make a significant contribution to the kinetic 
energy component $K_n$ of the Hamiltonian.  This is illustrated in
Fig.~(\ref{grad}).

Denoting by $\langle u_j \rangle$ and $\langle v_j \rangle$ the means 
of the variables $u_j$ and $v_j$ with respect to the yet to be
determined  ensemble $\rhon$, 
we now identify the coherent state with the mean--field 
pair $(\langle u^{(n)} (x) \rangle, \langle v^{(n)} (x) \rangle) 
= (\sum_{j=1}^n \langle u_j \rangle e_j(x),  
\sum_{j=1}^n \langle v_j \rangle e_j(x))$. The fluctuations, or 
small-scale radiation inherent in the long--time state then correspond 
to the difference  $(\delta \un, \delta \vn)  
\equiv (u^{(n)}-\unmean, v^{(n)}-\vnmean)$ between 
the state vector $(u^{(n)}, v^{(n)})$ and the mean--field vector. 
 The statistics of the fluctuations are encoded in the probability 
 density $\rhon$.  Based on the considerations of the preceding 
 paragraph, and the results of the numerical simulations displayed in 
Fig.~(\ref{grad}), it seems reasonable to conjecture that 
 the amplitude of the fluctuations of the field $\psin$ in the 
long-time state of the NLS system (\ref{spectrunc}) should vanish  
entirely (in some appropriate sense) in the continuum 
limit $n \rightarrow \infty$.  Thus we are led to the following 
{\em vanishing of fluctuations hypothesis}: 
 
\begin{equation} \label{vfh}  
\int_{\Omega} \left [ \langle (\delta \un )^2 \rangle +  
\langle (\delta \vn )^2\rangle \right ] dx \equiv 
\sum_{j=1}^n \left [ \langle (\delta u_j)^2 \rangle +   
\langle (\delta v_j)^2
\rangle \right ] 
\rightarrow 0\,,\; \mbox{as }\; n \rightarrow\infty\,. 
\end{equation} 
Here, $\delta u_j = u_j - \langle u_j \rangle$ represents the fluctuations 
of the Fourier coefficient $u_j$ about its mean value 
$\langle u_j \rangle$, and similarly for 
$\delta v_j$. We emphasize that (\ref{vfh}) is a hypothesis used to  
construct our statistical theory, and not a conclusion drawn from  
the theory itself.  
 
An immediate consequence of the vanishing of fluctuations hypothesis 
is that for $n$ sufficiently large, the expectation $\langle N_n \rangle$
of 
the particle number is determined almost entirely by the mean 
$(\unmean, \vnmean)$. 
Furthermore, the hypothesis (\ref{vfh}) implies that for $n$ large, 
the expectation $\langle \Theta_n(u^{(n)}, v^{(n)}) \rangle$ of the 
potential energy is well approximated by 
$\Theta_n(\unmean,\vnmean)$, which is the potential energy of the 
mean. This may be seen by expanding the potential $F$    
 about the mean 
$(\unmean,\vnmean)$ in equation (\ref{tn}), taking expectations, and
noting that because of the vanishing of fluctuations hypothesis 
  (\ref{vfh}), there holds $| \langle \Theta_n(u^{(n)}, v^{(n)}) \rangle -
\Theta_n(\unmean,\vnmean)| = o(1)$ as $n \rightarrow \infty$. 
 Notice, however, that the vanishing
of fluctuations hypothesis does not imply that the contribution of the  
fluctuations to the expectation of the kinetic energy becomes negligible
in the
limit $n \rightarrow \infty$.  Indeed, this contribution  
is $(1/2) 
\sum_{j=1}^n k_j^2 [ \langle (\delta u_j)^2 \rangle  
+\langle (\delta v_j)^2 \rangle ]$, which need not tend to 0 as $n 
\rightarrow \infty$, even if (\ref{vfh}) holds.  Thus, from these 
arguments,  we conclude 
that for $n$ sufficiently large, $\langle H_n \rangle \approx 
\frac{1}{2} 
\sum_{j=1}^n k_j^2( \langle u_j^2 \rangle + \langle v_j^2 \rangle ) 
- \frac{1}{2} \int_{\Omega} F(\unmean^2 + \vnmean^2)\, dx$. 
These considerations lead us to impose the following  
{\em mean--field constraints} on the admissible probability 
densities $\rho$ on the $2n$-dimensional phase space: 
 
\begin{equation} \label{mfc} 
\begin{array}{c} 
\tilde{N}_n(\rho) \equiv \frac{1}{2} \sum_{j=1}^n (\langle u_j \rangle^2 
+ \langle v_j \rangle^2) = N^0  \\ 
\\ 
\tilde{H}_n (\rho) \equiv \frac{1}{2} \sum_{j=1}^n k_j^2 (\langle 
u_j^2 \rangle + \langle v_j^2 \rangle)  
-\frac{1}{2} \int_{\Omega} F(\langle u^{(n)} \rangle^2 + \langle v^{(n)} 
\rangle^2)\,dx = H^0\,. \end{array} 
\end{equation} 
Here, $N^0$ and $H^0$ are the conserved values of the particle  
number and the Hamiltonian, as determined from initial conditions. 
The statistical equilibrium states are then  defined to be probability 
densities $\rhon$ on the phase--space ${\bf R}^{2n}$ that maximize the  
entropy (\ref{gbent}) subject to the constraints (\ref{mfc}).  
We shall refer to the constrained maximum entropy principle that  
determines the statistical equilibria as (MEP). 
 
Further justification and motivation for the vanishing 
of fluctuations hypothesis (\ref{vfh}), which leads to the  
mean--field constraints 
in the maximum entropy principle (MEP), are provided in \cite{JTZ}. 
In particular, it is proved in \cite{JTZ} that  the solutions $\rho^{(n)}$ 
of (MEP) concentrate 
on the phase--space manifold on which $H_n=H^0$ and $N_n=N^0$ in the
continuum limit  
$n \rightarrow \infty$, in the 
sense that $\langle N_n \rangle \rightarrow N^0$, 
$\langle H_n \rangle \rightarrow H^0$, and 
$ \mbox{var }N_n \rightarrow 0, \mbox{var }H_n \rightarrow 0$ in this 
limit. Here, $\mbox{var }W$ denotes the variance of the random 
variable $W$.  This 
concentration property establishes a form of asymptotic  
equivalence between the 
mean--field ensembles $\rho^{(n)}$ and the microcanonical ensemble, 
which is the invariant measure concentrated on the phase--space manifold
on which $H_n = H^0$ and $N_n=N^0$. It therefore provides a strong
theoretical  
justification for the  mean--field statistical model. 
 
\section{Calculation and Analysis of Equilibrium States} 
The solutions $\rhon$ 
of (MEP) are calculated by an application of the Lagrange 
multiplier rule 
\begin{displaymath}
 S'(\rhon) =  \mu  \tilde{N}'_n(\rhon) + \beta \tilde{H}'_n(\rhon)\,,
\end{displaymath} 
 where  $\beta$ and $\mu$ are the Lagrange 
multipliers to enforce that the probability density $\rhon$ satisfy
 the constraints  
(\ref{mfc}). A straightforward but tedious 
calculation yields the following expression for the maximum entropy 
distribution $\rho^{(n)}$\cite{JTZ}: 
 
\begin{equation}\label{rho1} 
\rho^{(n)}(u_1, \ldots, u_n, v_1, \ldots, v_n) = \prod_{j=1}^n \rho_j(u_j, 
v_j)\,, 
\end{equation} 
where, for $j=1,\ldots, n$,  
 
\begin{equation}\label{rho2} 
\rho_j(u_j,v_j) = \frac{\beta k_j^2}{2 \pi} \exp \left \{ -\frac{\beta 
k_j^2}{2}\left ((u_j - \langle u_j \rangle)^2 
+ (v_j - \langle v_j \rangle)^2 \right ) \right \}\,, 
\end{equation} 
with: 
\begin{equation}\label{uv} 
\begin{array}{c} 
\langle u_j \rangle = \frac{1}{k_j^2} \left ( f(\langle u^{(n)} 
\rangle^2 + \langle v^{(n)} \rangle^2) 
\langle u^{(n)} \rangle  \right )_j -\frac{\mu}{\beta k_j^2}\langle 
u_j \rangle  \\ 
\\ 
\langle v_j \rangle = \frac{1}{k_j^2} \left ( f(\langle u^{(n)} 
\rangle^2 + \langle v^{(n)} \rangle^2) 
\langle v^{(n)} \rangle  \right )_j -\frac{\mu}{\beta k_j^2} \langle 
v_j \rangle\,.  
\end{array}  
\end{equation} 
Thus, for each $j$, $u_j$ and $v_j$ are independent Gaussian variables,
with 
means given by the nonlinear equations (\ref{uv}) and with identical
variances

\begin{equation} \label{variance}
\mbox{var } u_j = \mbox{var } v_j =  \frac{1}{\beta k_j^2}\,.
\end{equation}
Note that var $u_j = \langle (\delta u_j)^2 \rangle$ by definition, 
and likewise for $v_j$.
Obviously, the multiplier $\beta$ must be 
positive. Notice also that, since the probability density 
$\rho^{(n)}$ factors according to (\ref{rho1}),  the Fourier 
modes $u_j, v_j, j=1,\cdots, n$,  are mutually uncorrelated. 
In addition, we see from (\ref{uv}) that 
the complex mean--field $\langle \psi^{(n)} \rangle =  
\unmean + i \vnmean$ is solution of (setting $\lambda= \mu/\beta$) 
\begin{equation}\label{rho} 
\langle \psi^{(n)} \rangle_{xx}+P^n \left (f(|\langle \psi^{(n)} 
  \rangle|^2)\langle \psi^{(n)} \rangle \right ) - \lambda \langle 
\psi^{(n)}\rangle = 0\,, 
\end{equation} 
which is clearly the spectral truncation of the eigenvalue equation 
(\ref{gse}) for the continuous NLS system (\ref{nls1}). It follows,
therefore, that the mean--field  
predicted by our theory corresponds to a solitary wave solution of  
the NLS equation. Alternatively, the mean $(\langle u^{(n)} 
\rangle, \langle v^{(n)}\rangle)$ is a solution of the variational 
equation $\delta H_n + \lambda \delta N_n  = 0$, where $\lambda$ 
is a Lagrange multiplier to enforce the particle number constraint  
$N_n=N^0$. 
 
Now, as the maximum entropy distribution $\rho^{(n)}$ 
is required to satisfy the mean--field Hamiltonian 
constraint (\ref{mfc}), it follows from (\ref{rho1})--(\ref{rho}) that 
\begin{equation} \label{hameqn1} 
H^0  = \frac{n}{\beta} + H_n ( \langle u^{(n)} \rangle, \langle v^{(n)} 
\rangle)\,. 
\end{equation} 
The term $n/\beta$ represents the contribution  
to the kinetic energy from the Gaussian fluctuations, and $ H_n ( \langle
u^{(n)} 
\rangle,\langle v^{(n)}\rangle)$ is the Hamiltonian of the mean.  
Notice that the contribution of the fluctuations to the kinetic  
energy is divided evenly among the $n$ Fourier modes. From 
(\ref{hameqn1}), we obtain the following expression for $\beta$ in terms
of 
the number of modes $n$ and the Hamiltonian  
$ H_n ( \langle u^{(n)} \rangle,\langle v^{(n)} \rangle)$ of the mean: 
 
\begin{equation} \label{betan} 
\beta = \frac{n}{H^0 - H_n ( \langle u^{(n)} \rangle,\langle v^{(n)} 
\rangle)}\,. 
\end{equation} 
 
Using equations (\ref{rho1})--(\ref{betan}), we may easily 
calculate the entropy of any solution $\rhon$ of (MEP). This yields, after 
some algebraic manipulations, that  
\begin{equation} \label{entmax} 
S(\rho^{(n)}) = C(n) + n \log \left ( \frac{L^2 [H^0 - H_n ( \langle 
u^{(n)} \rangle,\langle v^{(n)} \rangle)]}{n} \right ) \,. 
\end{equation} 
where $C(n) = n - \sum_{j=1}^n \log (j^2 \pi/2)$ depends 
only on the number of Fourier modes $n$. Clearly, the entropy 
 $ S(\rho^{(n)})$ will be maximum if and only if the mean field pair  
$(\langle u^{(n)} \rangle, \langle v^{(n)} \rangle)$ corresponding 
to $\rho^{(n)}$  realizes the minimum possible value of $H_n$ over all
fields 
$(u^{(n)},v^{(n)})$  that satisfy the  constraint  
$N_n(u^{(n)}, v^{(n)}) = N^0$.

Equation (\ref{entmax}) reveals that in statistical equilibrium 
 the entropy is, up to 
additive and multiplicative constants,  the logarithm of the 
the kinetic energy contained in the turbulent fluctuations about  
the mean state.  This result, therefore, provides a precise interpretation 
to the notions set forth by  Zakharov {\em et al.} \cite{ZPSY} and Pomeau 
\cite{Pomeau} that the entropy of the NLS system is directly related 
 to the amount of kinetic energy contained in the small-scale fluctuations, 
 and that it is ``thermodynamically advantageous'' for the solution of NLS 
 to approach a ground state which 
 minimizes the Hamiltonian for the given number of particles.  
 
We now know that $H_n ( \langle u^{(n)} \rangle,\langle 
v^{(n)} \rangle) = H_n^*$, 
where $H_n^*$ is the minimum vale of $H_n$ allowed by the particle number 
constraint $N_n = N^0$. As a consequence, the Lagrange multiplier 
$\beta$ is uniquely determined by (\ref{betan}): 
\begin{equation} \label{betan2} 
\beta = \frac{n}{H^0 - H_n^*}\,. 
\end{equation} 
That the ``inverse temperature'' $\beta$ scales linearly with the 
number of Fourier modes $n$ is required in order to obtain a 
meaningful 
continuum limit $n \rightarrow \infty$ in which the 
expectations of the Hamiltonian and particle number remain finite. 
The scaling of the inverse temperature 
with the number of modes is a common feature 
of the equilibrium statistical mechanics of finite dimensional
approximations
of  other plasma and fluid 
systems with infinitely many degrees of freedom, as well \cite{MWC}.
The parameter $\lambda$ (which depends on $n$) is  also  determined by 
the requirement that the mean $ (\langle \un \rangle, \langle \vn \rangle
)$ realize
the minimum value of the Hamiltonian $H_n$ 
given the particle number constraint $N_n=N^0$.     
 
Using eqns.~(\ref{variance}) and (\ref{betan2}), 
we may now obtain an exact expression for the 
contribution of the fluctuations to the expectation of the particle  
number. This is 
\begin{equation} \label{partfluct} 
\frac{1}{2} \sum_{j=1}^n \left [ \langle (\delta u_j)^2 \rangle + 
\langle (\delta v_j)^2 \rangle  \right ]  
 =  \frac{H^0 - H_n^*}{n} \sum_{j=1}^n \frac{1}{k_j^2}   
 =  O(\frac{1}{n})\,,\;\; \mbox{as } n \rightarrow \infty\,. 
\end{equation} 
Recall that in the derivation of the mean--field constraints  
(\ref{mfc}), we 
assumed the vanishing of fluctuations condition (\ref{vfh}).  The 
calculation (\ref{partfluct}) shows, therefore, that the maximum entropy 
distributions $\rhon$ indeed satisfy the hypothesis (\ref{vfh}), and 
hence, that  the mean--field statistical theory is consistent with 
the assumption that was made to derive it.  But as the analysis of 
this section has shown, the maximum entropy distributions $\rhon$ 
provide much more information than is contained in the hypothesis  
(\ref{vfh}).  Most importantly, we know that the mean--field 
corresponding to $\rhon$ is an absolute minimizer of the Hamiltonian 
$H_n$ subject to the particle number constraint $N_n = N^0$. 
In addition, the theory  yields predictions for the particle number  
and kinetic energy spectral densities, at least for the $2n$-dimensional 
 spectrally truncated NLS 
system (\ref{spectrunc}) with $n$ large. Indeed, we have 
the following prediction for the particle number spectral density 
\begin{equation} \label{partspectra}  
\langle |\psi_j|^2 \rangle =  | \langle \psi_j \rangle |^2
 + \frac{H^0 - H_n^*}{n k_j^2}\,,
\end{equation} 
where we have used the identity $\psi_j = u_j + i v_j$, and 
eqns.~(\ref{variance}) 
and (\ref{betan2}).
The first term on the right hand side of (\ref{partspectra})  
is the contribution to the particle number spectrum from the mean, and 
the second term is the contribution from the fluctuations. Since the mean
field is a smooth
solution of the ground-state equation, 
its spectrum decays rapidly, so that for $j>>1$, we have
the approximation 
$\langle |\psi_j|^2 \rangle  
\approx (H^0 - H_n^*)/(n k_j^2)$.  The kinetic 
energy spectral density is obtained simply by 
multiplying eqn.~(\ref{partspectra}) by $k_j^2$.  
As emphasized above, we have the prediction that the   
 kinetic energy arising from the fluctuations is equipartitioned  
among the $n$ spectral modes, with 
each mode contributing the amount $(H^0 - H_n^*)/n$. 

While we have chosen to present the 
statistical theory specifically for homogeneous Dirichlet
boundary conditions, it is straightforward to develop
the theory for NLS on a periodic interval of length $L$, as well. In this
case, it is most convenient to write the spectrally truncated complex
field $\psin$ as 
\begin{displaymath}
\psin= \sum_{j=-n/2}^{n/2} \psi_j \exp ( i k_j x)\,,
\end{displaymath}
for $n$ an even positive integer, where $k_j = 2 \pi j/L$. 
The predictions of the statistical theory remain the same
as in the case of Dirichlet boundary conditions. In particular,
the mean field $\langle \psin \rangle$ is a minimizer of the Hamiltonian
$H_n$ given the particle number constraint $N_n=N^0$, and
the particle number spectrum satisfies (\ref{partspectra}) for $j \neq 0$
The Fourier coefficient $\psi_0$ may be consistently 
chosen to be deterministic 
(i.e., $\mbox{var }\psi_0 = 0$ and $\langle\psi_0 \rangle \equiv \psi_0$),
to eliminate the ambiguity arising from the 0 mode..

\section{Numerical results} 
 
The general predictions of the statistical theory developed above
do not depend crucially on the 
particular nonlinearity $f$ in the NLS equation (\ref{nls1}). 
Indeed, for any $f$ satisfying the conditions stated in the introduction,
 the coherent structure
predicted by the theory in the continuum limit $n \rightarrow \infty$
corresponds to the 
solitary wave  
that minimizes the energy for the given number of particles $N^0$.
Also, for any such nonlinearity $f$,  the particle number spectrum in the
long-time limit for the spectrally truncated NLS system (\ref{spectrunc}),
according
to the statistical theory, should obey the relation
(\ref{partspectra}). Of course, the minimum value $H_n^*$ of the
Hamiltonian $H_n$ which enters this formula does depend on $f$.

Here, we will present numerical results primarily for periodic
boundary conditions and for the  
focusing power law nonlinearity $f(|\psi|^2) = |\psi|$. 
That is, we shall solve numerically the particular NLS equation

\begin{equation} \label{snls}
i \partial_t \psi + \partial_{xx} \psi + |\psi|\psi = 0\,,
\end{equation}
on a periodic interval of length $L$.
We have, however, carried out similar numerical experiments for
different focusing
nonlinearities and for Dirichlet boundary conditions, and we observed
that the general qualitative features of the long-time
dynamics are unaltered by such changes. 
The nonlinearity $f(|\psi|^2) = |\psi|$ actually
 represents a nice compromise between  
the focusing effect and  nonlinear interactions. 
For weaker  nonlinearities (such as the saturated ones), the 
interaction between modes is  
weak, and the time required to approach an asymptotic equilibrium
state is quite long.
On the other hand, for stronger nonlinearities, the solitary wave
structures that emerge exhibit narrow peaks of large amplitude, and  
therefore, greater spatial resolution is required in the numerical
simulations.  

The numerical scheme that we use for solving (\ref{snls}) is 
the well-known split-step Fourier method for a given number $n$ of Fourier
modes.  Throughout the duration of the simulations, 
the relative error in the particle number is kept at less than
 $10^{-6}$ percent, and the relative error in the Hamiltonian 
is no greater than $0.1$ percent. Notice that the numerical
simulations,  performed naturally 
for a finite number of modes, provide an ideal context 
for comparisons with  the mean--field statistical theory outlined above.

\begin{figure} 
\centerline{  \epsfxsize=10truecm \epsfbox{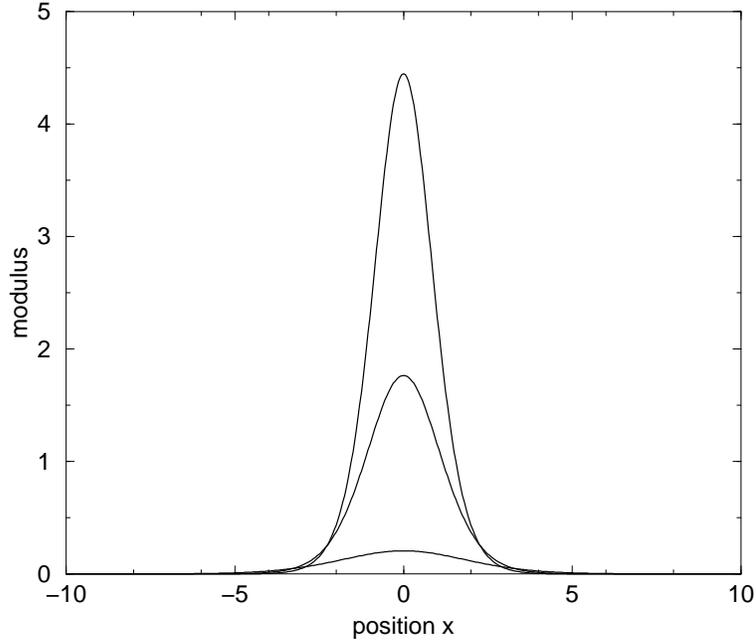}} 
\caption{\protect\small Solitary wave solutions of 
equation (\ref{snls}) for  particle numbers $N=1, 5$ and $10$. 
The modulus $|\psi|^2$ is shown as a function of the
position. 
\label{fsol}} 
\end{figure}

On the whole real line, the nonlinear Schr\"odinger equation
(\ref{snls}) has solitary wave solutions of the form
$ \psi(x,t)=\phi(x)e^{i\lambda^2 t}$, with  
\begin{equation} 
 \phi(x)=\frac{3\lambda^2}{2 {\rm cosh}^2(\frac{\lambda (x-x_0)}{2})}  
\label{solit} 
\end{equation} 
The particle number $N$ and the Hamiltonian $H$ of   
these soliton--like solutions are determined by 
the parameter $\lambda$ 
through the relationships $N=6 \lambda^3$ and 
$H=-\frac{18}{5} \lambda^5 $. These solutions are centered at 
$x=x_0$, as  shown in Figure (\ref{fsol}), and because of 
the focusing property
of equation (\ref{snls}), as $N$ increases, the amplitude of the
solitary wave increases, while its width
decreases. For a given value of the particle number $N$, 
the solitary wave (\ref{solit}) is the global
minimizer of the Hamiltonian  $H$ (when the integrals in the
definitions (\ref{ham1}) and (\ref{part1}) of the Hamiltonian and the
particle number extend over the real line).
Of course, the solitary wave solutions 
 for the equation (\ref{snls}) on a finite
interval,  
as well as those for the spectrally-truncated version  
(\ref{spectrunc}),  differ from the solution
(\ref{solit}) over the infinite interval. 
However, because the solitary waves (\ref{solit}) exhibit an
exponential decay,  for a large enough interval, 
and for a large enough number of
modes $n$,  such differences can be neglected for 
all practical purposes.

For constant $A$, the condensate 
$\psi(x,t)= A e^{i A t}$  is an
equilibrium solution of (\ref{snls}). However,
since the nonlinearity is focusing, this spatially homogeneous
solution is   
modulationally unstable. Indeed, if we expand $\psi$ around this  
solution in a series of the form 
 
$$ \psi(x,t)=(A + \sum \psi_k e^{(\sigma t+i kx)}) e^{i A t}$$ 
we obtain the dispersion relation: 

$$ \sigma^2= Ak^2-k^4 $$ 
Thus, the condensate is stable for $k^2 > A$, and
unstable for $k^2 < A$.  The most unstable
wave-number is $k_i=\sqrt{A/2}$. 

We choose to present in this paper
the following  set of numerical simulations: starting with the spatially
homogeneous solution $\psi(x,t=0)= A$ (with
$A$ of order $1$), we add initially a small  spatially
uncorrelated random  perturbation,  so that the modulational instability
develops.  Although we have checked that the long--time behavior of
the solution is not dependent on the initial conditions, except through
the initial and conserved
values $N^0$ and $H^0$ of the particle number and the Hamiltonian, 
this class of initial conditions is particularly convenient
for our purposes.  For example, by considering different realizations
of the initial random perturbation, we may
 perform  an ensemble average over different initial
conditions for a given $A$ (and therefore for fixed $N^0$ and
$H^0$).  Such initial conditions provide interesting analogies
to standard fluid turbulence problems,
 as we will emphasize in the conclusion.  

The spatially
uniform initial conditions we consider here may be thought of as being
 far away from the expected statistical attractor described by the
 maximum entropy probability density $\rhon$.  
Indeed, the spectrum of the condensate differs
considerably from the predicted statistical 
equilibrium spectrum (\ref{partspectra}).   The numerical
simulations that we perform here provide strong evidence that the
solutions of the spectrally truncated NLS system converge in the
long-time limit to a state that may be considered as statistically
steady. We shall compare the statistical properties of this long-time
state with the predictions of the mean--field statistical theory
that was developed and analyzed above.  
In addition, we shall also investigate the following questions
concerning the nature of the evolution leading from the initial state
to the long-time statistical equilibrium state:
1- How long does it take for the system to reach the vicinity of its 
statistical attractor, so that subsequently its statistical features 
may be considered as stationary?
2- How well can we characterize the ``path'' that a solution follows
en route to the statistically
steady state? That is, what are the generic features of the
transitory dynamics?

Figure (\ref{evolu}) demonstrates  that the transitory dynamics 
can be roughly decomposed into 
three stages: in the first stage, illustrated in Figure \ref{evolu} a),
 the modulational instability
creates an array of
soliton-like structures separated by a typical distance $l_i=2 \pi/k_i$   
associated with  most unstable wave number $k_i$. 
The second stage is characterized by the interaction and coalescence
of these solitons.  In this coarsening process, the number of solitons
decreases, while the amplitudes of the surviving solitons increase, 
until eventually a single soliton of large amplitude persists
amongst a sea of small-amplitude background  
radiation (Figures \ref{evolu} b) and c)).  This intermediate
stage has previously been observed for other nonlinear
Schr\"{o}dinger equations in one and two spatial dimensions 
\cite{ZPSY,SC},  and it was shown in \cite{SC} that this coarsening
process follows a self-similar dynamics. The dynamical exponents
of these processes are not very well understood at this point,
however.  During the final stage of the dynamics, the surviving
large-scale soliton interacts with the small-scale fluctuations. 
As time increases, the amplitude of the soliton
increases, while the amplitude of the fluctuations decreases 
(note the changes from Figure \ref{evolu}
 c) to Figure \ref{evolu} d)). In this stage of the dynamics, 
the mass (or number of
 particles) is gradually transferred from the small-scale fluctuations to
the large-scale coherent soliton. 
 For a finite number of modes $n$, the dynamics eventually reaches a
``stationary'' state whose properties are very well described by
 the mean--field statistical equilibrium theory developed above, as we
 shall demonstrate.
This  implies that long-time state may, in fact, be
thought of as a ``statistical attractor'', in the sense that,
according to the statistical theory, 
it corresponds to a  maximizer of the entropy
functional (\ref{gbent}) subject to the dynamical constraints (\ref{mfc}). 
Note that because the dynamics is reversible, 
intermediate states such as those in Figure (\ref{evolu} b)  
theoretically could still be attained even after the statistical
equilibrium state has been reached. In fact, a numerical simulation starting
from the state in Fig.~(\ref{evolu} d))
but with
the time step taken negative shows the reverse dynamics 
up to round-off errors, where
one can observe the decomposition of the solution into an array of
soliton-like
structures as in figure (\ref{evolu} a)) for intermediate times, while in
the limit $t \rightarrow -\infty$ an equilibrium state such as the one
of figure (\ref{evolu} d)) is once again attained. 

\begin{figure} 
\centerline{  \epsfxsize=10truecm \epsfbox{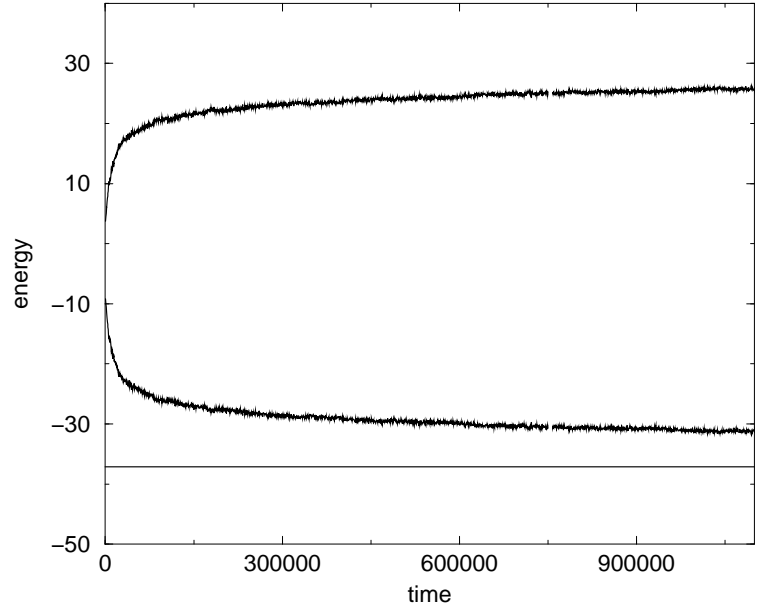}} 
\caption{\protect\small Time evolution of the kinetic (upper curve) and  
the potential (middle curve) energies. 
The kinetic energy is increasing and consequently the  
potential energy is decreasing, {\it in accord} with the statistical
theory 
developed above. The lower line indicates the potential  
energy of the solitary wave that contains all the particles 
of the system. The curves are obtained from an ensemble average over 
$16$ initial conditions for $n=512$. The length of the system is  
$L=128$, and the (conserved) values of
the particle number and the Hamiltonian are, respectively,
 $N^0=20.48$ and $H^0=-5.46$. 
\label{poten}} 
\end{figure}

The tendency of the solution of the NLS system (\ref{snls}) to approach the 
statistical equilibrium state is also captured in the evolution of the
kinetic and potential energies (see Figure 
\ref{poten}). While the sum of these two quantities, which is the
Hamiltonian, remains constant in
time, we observe that the kinetic energy increases monotonically,
and, consequently,  the potential energy decreases monotonically
as time goes on.  The initial time period where these quantities
evolve most rapidly (say $t < 20000$) corresponds to the first two
stages of the dynamics described above, in which the modulational
instability
creates an array of soliton-like structures which then coalesce into a
single coherent soliton.
After the coalescence has ended, the kinetic (potential) energy 
increases (decreases) very slowly to its saturation value.  In the
process, fluctuations develop on finer and finer spatial scales, which
accounts for the gradual increase of kinetic energy, while the
surviving soliton slowly absorbs mass from the background
fluctuations, thereby increasing the magnitude of the contribution to
the potential energy from the coherent structure.  In the long-time
limit, therefore, the
soliton accounts for the vast majority of the potential energy,  
while the fluctuations make a substantial contribution to the kinetic
energy.

\begin{figure} 
\centerline{  \epsfxsize=10truecm \epsfbox{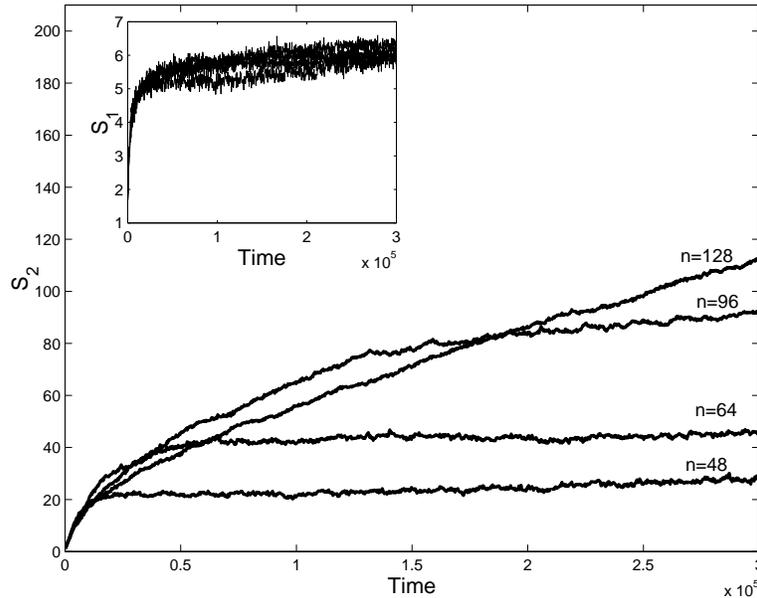}} 
\caption{\protect\small $S_2$ as a function of time for $n=48,64,96$ and  
$128$ (lower to upper curve). The simulations are all performed for  
a box of size $L=38.4$, for $N^0=9.6$ and $H^0=-3.2$. The curves  
are obtained from an ensemble average over $16$ runs for each $n$.  
Saturation is reached for $n=48$ and $n=64$, while it is  
almost obtained for $n=96$. As $n$ is increased,  the time required to
reach saturation increases, and the saturation value also increases.   
The inset
shows the kinetic energy $S_1$ as a function of time for the 
same values of $n$. 
As opposed to $S_2$, the saturation of $S_1$ seems to be occurring at 
about the same rate
for each $n$.  The saturation value of $S_1$  increases slightly as $n$
increases, but it remains bounded above by the statistical equilibrium
value $\langle S_1 \rangle$.
\label{satur}} 
\end{figure}

The mean--filed statistical theory 
provides a prediction for the expected value of
the kinetic energy $K_n$ in statistical equilibrium for a given number of
modes $n$.  This is $\langle K_n \rangle =  K_n(\langle \psin \rangle) + 
H^0 - H_n^*$, which follows directly upon multiplying 
eqn.~(\ref{partspectra}) by $k_j^2$ and summing over $j$. 
The first term  in this expression for $\langle K_n \rangle$ 
is the contribution to the  mean kinetic
energy from the coherent soliton structure which minimizes the
Hamiltonian $H_n$ subject to the particle number constraint $N_n=N^0$.
The second term in $\langle K_n \rangle$ is the contribution to the 
expectation of the kinetic
energy from the fluctuations. $H_n^*$ is the minimum value of $H_n$ given the
particle number constraint.
As $n \rightarrow \infty$, we see that $\langle K_n \rangle$ converges
to $K(\psi^\infty) + H^0 - H^*$, where $\psi^\infty$ is the minimizer of
the
Hamiltonian $H$ given the particle number constraint $N=N^0$ for 
continuous NLS system on the interval $[0,L]$, and  $H^* = H(\psi^\infty)$.  
Approximating $K(\psi^\infty)$ and $H(\psi^\infty)$ by $K(\phi)$ and
$H(\phi)$,
where $\phi$ is the solitary wave on the real line whose particle
number is $N^0$, we obtain for the setting considered in 
Figure (\ref{poten}) the large $n$ estimates $K_n(\langle \psin \rangle)
\approx 9.2, H^0-H_n^* \approx 22.4$, and therefore, $ \langle K_n
\rangle \approx 31.6$.  Also, according
to the statistical theory, the expected value $\langle \Theta_n
\rangle$ of the potential energy in statistical equilibrium should
converge as $n \rightarrow \infty$ to $\Theta(\psi^\infty)$.
Approximating
this by $\Theta(\phi)$, with $\phi$ as above, we have the estimate
$\langle \Theta_n \rangle \approx -37.1$, which we expect to be
accurate for sufficiently large $n$.  We see that
the kinetic (potential) energy of the numerical solution 
 is bounded above (below) by the estimate based on the
statistical theory, but as expected, the solution does not attain the
theoretically predicted value for a finite number of modes.
This is because, for the spectrally truncated system,
a finite amount of the particle number and the
potential
energy integrals are actually contained in the small-scale
fluctuations (according to the statistical theory, the contribution
of the fluctuations to these quantities should be $O(1/n)$, where $n$
is the number of spectral modes --this follows from
(\ref{partspectra}) \cite{JTZ}).  It may be checked that
the spatial resolution is improved (i.e., when the number of modes $n$ is
increased, while the length $L$ of the spatial interval, and the
values $H^0$ and $N^0$ of the Hamiltonian and the particle number are
held
fixed), the contributions of the fluctuations to the particle number
and the potential energy decrease, and the saturation values of the
kinetic and potential energy attained in the numerical simulations
come closer to the predicted statistical equilibrium 
averages of these quantities (see the inset in Fig.~(\ref{satur}),
which shows that the saturation value of kinetic energy increases
towards the predicted statistical equilibrium value
of modes $n$ in the numerical simulation increases). 
We expect that the contributions of the fluctuations to the particle
number and the potential energy should vanish entirely as $ n \rightarrow
\infty$ for fixed $L$, $H^0$ and $N^0$, and that
the predicted statistical equilibrium values for the mean kinetic
energy and potential energy should be approached very closely by
the numerical solution in the long-time limit when the number
of modes in the simulation is sufficiently large.

\begin{figure} 
\centerline{  \epsfxsize=10truecm \epsfbox{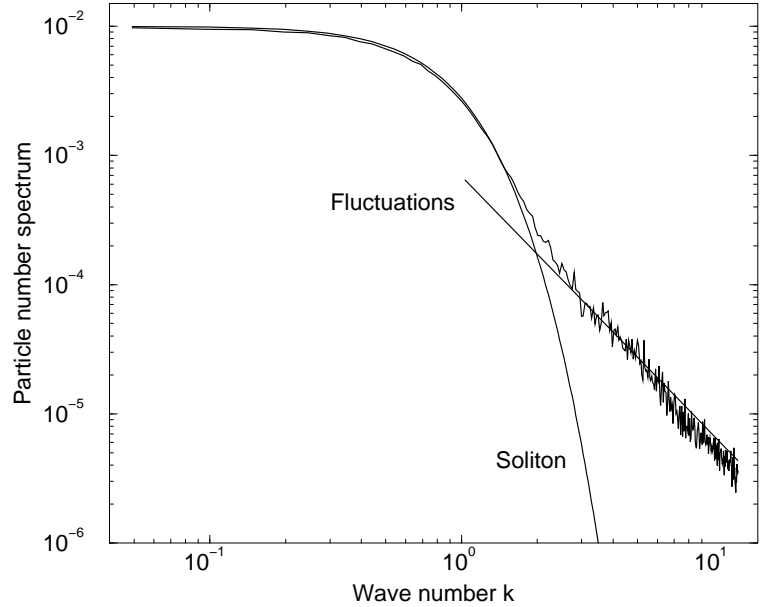}} 
\caption{\protect\small Particle number spectral density
 $ |\psi_k|^2$ as a 
function of $k$ for $t=1.1\times10^6$ 
unit time (upper curve). The lower curve  
(smooth one) is the particle number spectral
density  for the solitary wave  that
contains all the particles of the 
system. The straight line drawn for large $k$ corresponds to
the statistical prediction (\ref{partspectra}) for
the spectral density for large wavenumbers. The numerical simulation has 
been performed with $n=512$, $dx=0.25$, $N^0=20.48$ and $H^0=-5.46$. 
\label{spec}} 
\end{figure}

Figures (\ref{evolu}) and (\ref{poten}) clearly illustrate
that for a given (large) number of modes $n$, 
the dynamics converges when $t
\rightarrow \infty$ to a 
 state consisting of a large-scale coherent soliton, which accounts
for all but a small fraction  of
the  particle number and the potential
energy integrals, coupled with small-scale radiation, or fluctuations,
which account for  
the kinetic energy that is not contained in the coherent structure. 
Formula
(\ref{partspectra}) suggests, in fact, that in the long--time
limit, the coherent structure and the
background radiation exist in balance (or in statistical equilibrium) with
 each other, through the equipartition of kinetic
energy of the fluctuations. In Figure \ref{spec}),
we display the particle number spectral density $|\psi_k|^2$, where
$\psi_k$ is the Fourier transform of the field $\psi$,
as a function of the wave number $k$ for a long time run.
To obtain this spectrum, we have performed both an ensemble average
over 16 initial conditions, and a time average over the final 1000 time units
for each run.
For comparison, we have displayed in this figure  the spectrum of the
solitary wave (\ref{solit})  whose particle number is equal to
conserved value of the  particle number for the simulation.
Observe that there is both a qualitative and quantitative agreement
between the spectrum of this solitary wave solution and
the small wavenumber portion  of the spectrum arising from the numerical
simulations. This is in accord with the statistical equilibrium
theory, which
predicts that the coherent structure should coincide with this
solitary wave (in the limit $n \rightarrow \infty$).
 For larger wavenumbers, the spectrum of the numerical
solution is dominated by the small
scale fluctuations. We have indicated on the graph 
 the large wavenumber spectrum predicted
by the statistical theory. This prediction comes from the second
expression
on the right hand side of 
 eqn.~(\ref{partspectra}), except that we have approximated the
minimum
value  $H_n^*$ of the Hamiltonian for the spectrally truncated system
with $n$ modes by the Hamiltonian $H^*$ of the above-mentioned
solitary wave solution for the continuum system. 
Not only is there a good qualitative agreement with the predicted
equipartition of kinetic energy amongst the small-scale fluctuations
(i.e., the $k^{-2}$ slope),  but there is also an excellent 
quantitative agreement between the numerical results and the formula
(\ref{partspectra}) for large $k$. Let us mention that the 
long-time spectrum obtained from a single simulation starting
from a given initial
condition, and without time averaging, though similar to the spectrum
displayed in
Figure (\ref{spec}), is much noisier.

As we have mentioned above, the numerical spectrum shown in Figure
(\ref{spec}) arises from an ensemble average 
over long time and over different initial conditions (with the same
values
of the particle number and the Hamiltonian). Now, 
under the assumption that the dynamics is ergodic, 
such an average should
coincide with an average with respect to the 
microcanonical ensemble for the spectrally
truncated NLS system \cite{Balescu}.  
Since it can be shown that
the the mean--field statistical ensembles $\rhon$ constructed above
concentrate
on the microcanonical ensemble in the continuum limit $n \rightarrow
\infty$ (see Theorem 3 of reference \cite{JTZ}), 
it should be that averages with respect to
$\rhon$ for large $n$ agree with the ensemble average of the
numerical simulations over initial conditions and time, assuming
ergodicity of the dynamics.  While we have not shown that the
dynamics is ergodic, we have, in fact, demonstrated what we believe to
be a convincing agreement between the predictions of the
 mean--field ensembles $\rhon$
and the results of direct numerical simulations.

We have also monitored the time evolution of the quantities

\begin{equation} \label{sm} 
S_m(\psin)= \sum_j k_j^{2m} |\psi_j|^2\,,
\end{equation}
for $m$ a positive integer.  For the periodic boundary conditions 
considered here,
 the index $j$ ranges from $-n/2$ to $n/2$ and the wavenumber $k_j$ is
given by
$k_j = 2 \pi j /L$. 
Note that $S_1$ is the kinetic energy. In general,
$S_m(\psi)$ is the squared $L_2$ norm of the m--th derivative 
of the field $\psi$.
The growth of $S_m$ in time is an indicator of the development 
of fluctuations of the field on fine spatial scales. 
In addition, we may consider that $S_m$ gives an estimate of 
the evolution of the typical wave number $K(t)$ of the fluctuations 
since, roughly speaking, we can estimate 
 $S_m \sim K(t)^{2(m-1)}$. 

The mean--field statistical theory provides the following prediction
for the expectation of $S_m$ in statistical equilibrium for a given number
of modes $n$:

\begin{equation} \label{smmean}
\langle S_m \rangle 
 =  \sum_{j= -n/2}^{n/2} k_j^{2m} |\langle \psi_j \rangle |^2
+ \left ( \frac{2 \pi}{L} \right )^{2(m-1)}
\frac{H^0-H_n^*}{n} \sum_{j=-n/2}^{n/2} j^{2(m-1)}\,,
\end{equation}
where we have used eqn.~(\ref{partspectra}).  
The first term is the contribution
to $\langle S_m \rangle$ from the mean field (the coherent structure), 
and the second term
is the contribution from the fluctuations.  
Note that for a finite number of modes
$n$, $\langle S_m \rangle$ is finite for each $m$, but 
 only  $\langle S_1 \rangle$, which is
the mean of the kinetic energy,  remains
finite in  the continuum limit $n \rightarrow \infty$.  The divergence
of $\langle S_m \rangle$ for $m \ge 2$ comes from the second expression
on the right hand side of (\ref{smmean}) (i.e., from the
fluctuations), 
which is of the order $n^{2(m-1)}$ as $n \rightarrow \infty$.  
For example, when $m=2$
this expression is found to be 
$ \pi^2 (H^0 - H_n^*)(n^2 + 3n + 2)/(3L^2)$, and we have the following
formula for $\langle S_2 \rangle$ for a given number of modes $n$ in
the spectrally truncated NLS system:

\begin{equation} \label{s2sat}
\langle S_2 \rangle =
 \sum_{j= -n/2}^{n/2} k_j^{2} |\langle \psi_j \rangle |^2  +
 \frac{ \pi^2 (H^0 - H_n^*)(n^2 + 3n + 2)}{3L^2}\,.
\end{equation}

Based on the considerations of the previous paragraph,
 we expect that the numerical simulations 
for given number of modes $n$ will reveal that the
quantities
$S_m$ are bounded and saturate when $t \rightarrow \infty$, 
but that the larger the number of modes $n$, 
the larger the saturation value of $S_m$
(at least for $m \ge 2$).
Figure (\ref{satur}) 
shows the evolution in time of $S_2$ for different values of $n$
(with the same $L$, $N^0$ and $H^0$).  
We observe that saturation does indeed occur for a 
finite number of modes. Also,  
as $n$ increases, the saturation value increases, as does the time
required to reach saturation. By approximating the sum in
 eqn.~(\ref{s2sat}) by $\int_{-\infty}^{\infty} |\phi_{xx}|^2\, dx$ and
 approximating
$H_n^*$ by $H(\phi)$, where $\phi$ is the solitary wave on the whole
 real line whose particle number  is equal to the conserved particle
 number for the simulations treated in Figure (\ref{satur}), we obtain
 the following estimates: $\langle S_2 \rangle \approx 27.1, 45.5,
 97.6$ and $170.3$ for $n=48, 64, 96$, and $128$, respectively.
Note that these estimates for $\langle S_2 \rangle$ agree closely with
the observed saturation values of $S_2$ in the numerical simulations
 for $n=48$ and $n=64$.  For $n=96$, saturation has not quite yet been
 reached, but the value of $S_2$ at the final time $t = 3 \times
 10^5$ is still reasonably close to the theoretical estimate of
 $97.6$. For $n=128$, $S_2$ is still growing considerably at the
 final time of the simulation, and so we can not make comparisons with
 the statistical prediction for $\langle S_2 \rangle$ at this point.
 The inset in Figure (\ref{satur}) shows the evolution of
the kinetic energy $S_1$ as a function of time for $n=48,64,96$, and
$128$.
We see that the kinetic energy saturates nearly at the 
same rate for all of the values
of $n$ considered here. Clearly, $S_1$ remains bounded
as $n$ increases.  As discussed above,   $\langle
S_1 \rangle$, the statistical equilibrium
value of of the mean kinetic energy, converges as $n \rightarrow
 \infty$ to $K(\psi^\infty) + H^0 -H(\psi^\infty)$, where $\psi^\infty$ is the
solitary
 wave that minimizes the Hamiltonian $H$ for the given particle number
 $N^0$ for the NLS system on the interval $[0,L]$.  Once again,
 approximating $\psi^\infty$ by the solitary wave $\phi$ on the
 whole real line that minimizes $H$ given the particle number
 constraint $N=N^0$, we may estimate the limiting value
$\langle S_1 \rangle$ by $K(\phi) + H^0 - H(\phi)$, which is
$\approx 7.3$ for the value $N^0 = 9.6$ considered in Figure
 (\ref{satur}).  This estimate provides an upper bound on the
 saturation values of $S_1$ observed in the simulations, and as
the number of modes in the simulations increases, $S_1$ saturates
 closer to this approximation of the statistical equilibrium value.

\begin{figure} 
\centerline{  \epsfxsize=10truecm \epsfbox{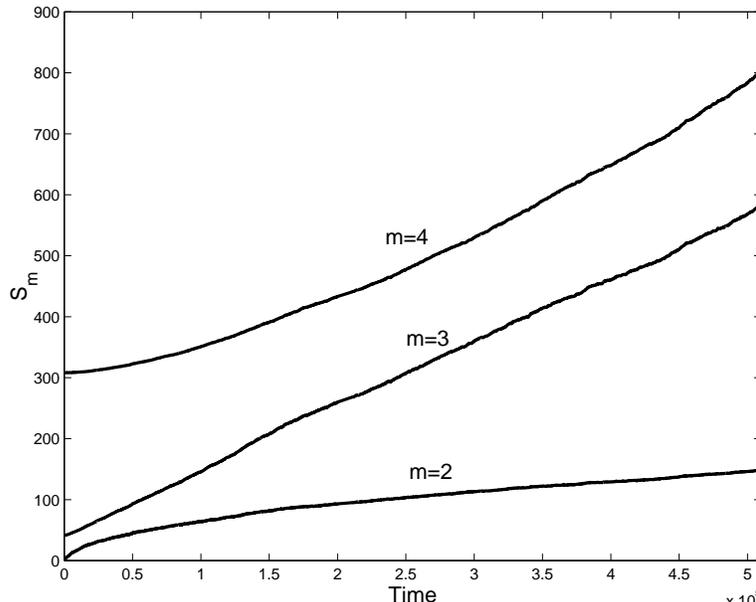}} 
\caption{\protect\small $S_m$ as a function of time for $m=2,3$ and  
$4$ (lower to upper curve). The  growth of these quantities is
indicative of the development of fluctuations on fine spatial scales
as time increases.   
The $S_m$ have been calculated for $512$ modes with  
$N^0=12.8$, $H^0=-4.26$ and $dx=0.1$, with an ensemble average  
over $16$ runs. They have been rescaled in order 
to display them on the same graph. 
\label{moment}} 
\end{figure}

When the spatial resolution of the numerical simulations is improved
(i.e, when $n$ is increased with $L$ fixed), 
the functions $S_m$ are typically seen to exhibit power law growth in time 
before reaching saturation (see Figure \ref{moment}).

Indeed, we observe for $m=2,3$ and $4$ that  $S_m$  obeys the 
following power law dynamics:

\begin{equation} \label{power}
 S_m \propto t^{2(m-1) \nu}\,,
\end{equation}
with $ \nu =0.25 \pm 0.01 $.
This behavior is  observed for $t$ large enough that the
coalescence process has ended. It corresponds, therefore, to the regime
where the kinetic energy has essentially reached saturation.  Remarkably,  
the observed dynamical exponent $\nu$ is in good agreement with the 
prediction of Pomeau \cite{Pomeau}, which estimates 
the evolution of the typical wave number
$K(t)$ of the fluctuations as time increases. The estimate comes from a
dimensional analysis of the weak turbulence equation deduced from
equation (\ref{snls}). Describing the fluctuation field $\delta
\psi$ as: 
$$ \delta \psi =\frac{1}{\sqrt{L}}\int dk (\delta I_k)^{1/2} 
e^{i(kx-\omega t)} , $$
the relation between the energy $\omega$ and the wave number $k$ is called
the spectrum of excitations (we refer the reader to \cite{Pomeau} for details).
In \cite{Pomeau}, it has been shown
that if this wave number $K(t)$ is in the range where the spectrum of 
excitations obeys $ \omega(k) \sim k^2$, which means that the 
fluctuations behave essentially like free particles, then, assuming 
that there is
two--wave resonance in the weak turbulence approximation, it follows
that $ K(t) \sim (\epsilon t) ^{1/4} $. Here, $\epsilon$ is
the  spatial energy density of the fluctuations (so $\epsilon \sim 
(H^0-H^*)/L$).  The analysis in
\cite{Pomeau} was carried out for cubic defocusing NLS in two spatial
dimensions, and does not immediately go over to the case 
under consideration here. In fact, strictly speaking, 
in the regime $\omega(k)= k^2$,
the resonance of two waves cannot hold in one spatial dimension. 
However, we conjecture that, 
due to the interactions with the large--scale coherent
structure, the resonance may in fact be meaningful in the present
setting, 
and therefore, we believe 
that a dimensional weak turbulence analysis along the lines 
of that developed in \cite{Pomeau} may be
relevant.  We hope to explore this possibility in the future.
Interestingly, for the NLS system, the approximation $\omega(k)= k^2$
is  usually valid in the limit $k \gg 1$. In the numerical simulations,
the finite number of modes provides an ultraviolet cutoff since the 
largest wave number of the system is
$$ k_{max}=\frac{\pi}{dx}=n \frac{\pi}{L} .$$
We remark that we have been able to recognize the power-law growth in
time of the quantities $S_m$ only for the smallest $dx$ we have 
considered in our simulations.  For larger  $dx$ the free particle regime 
might not
be realized, and it is not surprising in this case
 that the power law behavior is not
observed.

\begin{figure} 
\centerline{  \epsfxsize=10truecm \epsfbox{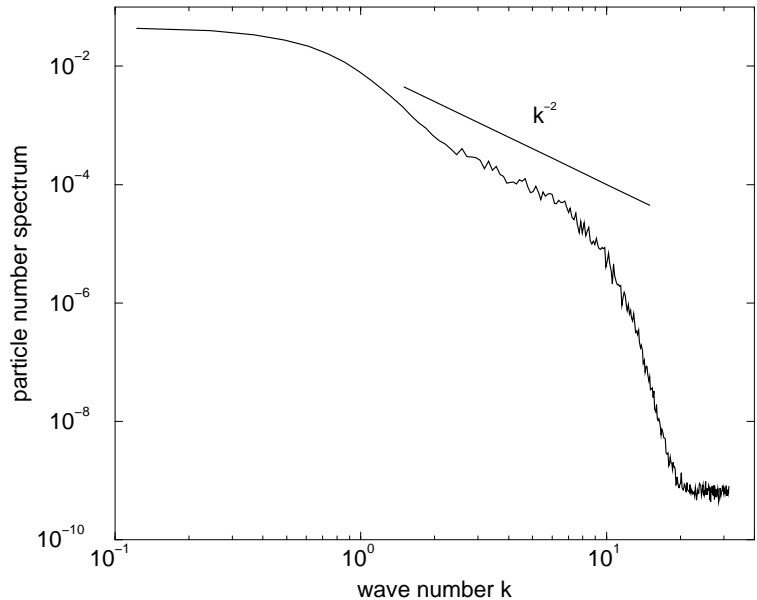}} 
\caption{\protect\small The particle number spectral density for
  $n=512$
and
$dx=0.1$
(thus $L=51.2$) at unit time $t=5 \cdot 10^5$. The coherent soliton
structure
 already accounts for almost the entire  number of particles
 of the system,  but 
the system has not yet reached  statistical equilibrium. The initial
noise level is still present for large wavenumbers ($ k \ge 20$), while 
at smaller wavenumbers, one can recognize both the soliton-like 
structure and a fluctuation spectrum following approximately a $k^{-2}$
law.
The spectrum has been obtained by an ensemble average over 16 initial 
conditions and a time average over the final 10 unit times.
\label{inter}} 
\end{figure}

The previous considerations allow us to attach a more precise meaning
to what we have been referring to as the transitory dynamics and the 
equilibrium state for the spectrally truncated
NLS system. For sufficiently small grid sizes $dx$,
 one may consider that  $S_m$ grows according to the power law
(\ref{power})  until the time $t_n$ at which the typical 
wave number $K(t)$ reaches the largest available wave number  $k_{max}$.
This time $t_n$  (which  appears to scale as 
$n^4$ for fixed $L$, $N^0$ and $H^0$) 
defines a crossover between the transitory regime in which the solution 
evolves towards the spectrum (\ref{partspectra}), and the statistical
equilibrium
regime where the system investigates its phase-space according 
to the  probability density $\rho^{(n)}$. Notice that in the continuum
limit $n \rightarrow \infty$, $t_n$ diverges to infinity, 
so that continuous NLS system can not reach 
statistical equilibrium in finite time.
Such conjectures  are supported by the investigation of the 
dynamics of the particle number spectrum during the intermediate 
time regime after which the coarsening process has ended, 
but before the final
statistical equilibrium state has been reached. Note that the statistical
equilibrium model does not provide predictions about the time evolution
of the spectrum, because it is strictly an equilibrium theory.
In fact, based on the statistical theory alone, nothing
can be said about  path which the system follows from the
statistically unlikely initial condition to the final statistical 
equilibrium state.
Figure (\ref{inter}) displays the particle number spectrum at
$t=5 \cdot 10^5$ unit time  for a spatial resolution of $dx=0.1$.

 This figure illustrates that the system investigates smaller and
smaller scales as time increases. Indeed,  the particle number
 spectrum for the modes $k \ge 20$ is still at the level of the
initial noise. Thus, the smallest scales available to the system
have yet to be excited at the time $t=5\cdot 10^5$.
For larger scales, however,  one can recognize both the spectrum
corresponding to the coherent soliton structure, 
 and the fluctuation spectrum which
appears to follow, at least approximately, 
the equilibrium law $ |\psi_k|^2 \propto k^{-2} $.
This suggests the following scenario for the spectrally truncated NLS 
dynamics: as time
increases, smaller and smaller scales are explored, 
until eventually all available modes are excited.
However, at any given time after the coalescence process has ended, 
the system may be considered as being in
statitistical equilibrium
 over all the modes that have been excited up
to that point.
Defining $k_{max}(t)$ to be the largest wave number 
that the system has
reached up to time $t$, then we obtain from (\ref{power}) 
that $k_{max}(t)
\sim (\epsilon t)^{1/4}$, if $k_{max}$ is large enough. Now if we
denote by $n(t)$ the number of modes that have been excited
up to time $t$, then based on our previous arguments, we have that
$k_{max}(t) = \pi n(t)/L$.  But, using $k_{max}(t)
\sim (\epsilon t)^{1/4}$,
 we obtain the following estimate for the 
spectrum of the fluctuations at time $t$:

\begin{equation} \label{fluctspect} 
\langle |\delta \psi_k|^2 \rangle \sim \frac{H^0-H^*}{n(t) k^2} \sim \frac{\pi
\epsilon^{3/4}}
{t^{1/4} k^2}\,,
\end{equation}
for $|k| \le k_{\max}(t)$.
The total particle number spectrum at time $t$, of course, 
has to be taken as the sum of the spectrum corresponding to the 
large-scale coherent structure (which decreases exponentially for large k) 
and the spectrum of the fluctuations.

We emphasize that the derivation
of the equation (\ref{fluctspect}) describing the particle number spectrum  
at an intermediate time $t$ crucially depends on the assumption
that the system evolves in such a way that it is
nearly in statistical equilibrium
over all the modes that have been
investigated up to that time.  The Figure (\ref{inter}) has motivated
us to make this assumption, but clearly further numerical
investigations should be carried out in order to test the validity of 
this hypothesis, as well as the accuracy of the 
formula (\ref{fluctspect}).  Nevertheless, we find it quite
interesting that the  fluctuation spectrum
(\ref{fluctspect}) agrees with the prediction in \cite{Pomeau}, which
was derived from a dimensional analysis of the weak turbulence
equations for the NLS system.

\section{Discussion-Conclusion-Acknowledgements.}

The primary purpose of the present work has been to test the
predictions of a mean--field statistical model of self-organization
in a generic class of nonintegrable focusing NLS equations defined by 
eqn.~(\ref{nls1}).  This statistical theory, which has been summarized
above, was originally developed and analyzed in \cite{JTZ}.  
In fact, we have demonstrated a remarkable 
 agreement between the predictions of the statistical
theory and  the results of  direct numerical
simulations of the NLS system.  There is a
strong qualitative and quantitative agreement between the 
mean field predicted by the statistical theory and the large-scale
coherent structure observed in the long-time numerical simulations.
In addition, the statistical model accurately predicts the
the long-time spectrum of the numerical solution of the NLS system.
The main conclusions we have reached are 1) The coherent structure
that emerges in the asymptotic time limit is the solitary wave that
 minimizes  the system Hamiltonian subject to the 
particle number constraint $N=N^0$, 
where $N^0$ is the given (conserved) value of $N$, and  2)  The
difference between the conserved Hamiltonian and the Hamiltonian
of the coherent state resides in Gaussian 
fluctuations equipartitioned over wavenumbers.

While the statistical model we have developed is an equilibrium 
theory, and, strictly speaking, only provides predictions concerning
the long-time  statistical properties of the NLS system, we have
combined this theory with insight gained from numerical
simulations to paint a picture of the nature of the dynamics leading
up to the statistical equilibrium state.  Specifically, the 
simulations 
(and, in particular, the results shown in Fig.~\ref{inter}) indicate
that evolution is  such  that, at a given time after
the coarsening process has ended, the system is nearly in 
statistical equilibrium over all the modes that have been excited by
that time. 
From this observation the fact that the quantities
$S_m$ defined in (\ref{sm}) are seen to
 exhibit the power law growth
in time  according to eqn.~ (\ref{power}), we have arrived
at the prediction (\ref{fluctspect}) for the time--dependence of
the spectrum of the fluctuations.  As we have mentioned above, 
 results such as (\ref{power}) and (\ref{fluctspect}) have
 previously been derived
by Pomeau \cite{Pomeau}
from weak turbulence arguments, but for the defocusing cubic
NLS equation in a bounded two-dimensional spatial domain.  
We believe that it would be an interesting exercise to check whether
these formulas  can be derived
directly from a weak turbulence analysis in the present context --that
is, for 1D nonintegrable focusing NLS equations in the absence
of collapse.

We would like to point
out certain analogies  between the dynamics of
the NLS systems we have considered here, and the dynamics of a
turbulent 2D Navier-Stokes fluid.  A prominent feature of large Reynolds
number 2D Navier-Stokes turbulence is the formation of quasisteady 
coherent vortex structures \cite{tab,vor,Mcwilliams}. Starting from
generic initial conditions, the evolution of the fluid is
characterized by the formation of a collection of large-scale
vortices, and the subsequent merger or coalescence of like-signed
vortices \cite{Mcwilliams}.  The large-scale soliton structures 
in (focusing, nonintegrable) NLS play a role similar to that of the
vortices in 2D Navier-Stokes turbulence.  Indeed, we have observed in our
numerical simulations of NLS the formation of an array of 
soliton-like structures which eventually coalesce into a single
persistent soliton of large amplitude.  Another characteristic feature
of turbulence in two-dimensions is the presence of a dual 
cascade \cite{krai}.  There is a direct cascade of enstrophy to small
scales and an inverse cascade of energy to large scales.  
As pointed out long ago by Kraichnan
\cite{krai}, the existence of the inverse cascade of energy is 
indicative of the formation of a large-scale structure in the system.
In NLS, there is also a dual cascade. Indeed, our numerical simulations, 
which correspond to
injecting as initial conditions particle number and energy at a given
scale $l_i$ associated with the modulational instability,
have revealed that their is a direct transfer of kinetic
energy to spatial scales smaller than $l_i$, 
while the particle number is transferred to
large scales.  While the 1D NLS equation 
is much simpler
system to investigate, both 
analytically and numerically, than a turbulent 2D fluid system,
we believe that the understanding of
the coalescence and transfer processes in this generic model
of nonlinear wave turbulence  might provide important
insight into the nature of turbulent systems in general.

It is a pleasure to thank Robert
Almgren, Shiyi Chen, Leo Kadanoff, Yves Pomeau,
Bruce Turkington, and Scott Zoldi
for valuable discussions and
suggestions. R.~J. acknowledges support from an NSF
Mathematical Sciences Postdoctoral Research Fellowship and from the
DOE through a grant to the Center for Nonlinear Studies at Los Alamos
National Laboratory. 
The research of C.~J. has been supported by 
the ASCI Flash Center at the University of Chicago 
under DOE contract B341495.

\end{document}